\documentstyle[11pt,iau185,twoside,epsf]{article}

\begin{document}
\title{Pulsating AGB stars in the LMC}
\author{Jacco Th.\ van Loon}
\affil{Astrophysics Group, School of Chemistry \& Physics, Keele University,
Staffordshire ST5 5BG, United Kingdom (jacco@astro.keele.ac.uk)}

\begin{abstract}
I give a brief review and interpretation of the evolution, mass loss and
pulsation of AGB stars in the Large Magellanic Cloud.
\end{abstract}


\section{The Magellanic Clouds in an astro-physical context}

Our view of the place of mankind in the Universe changed dramatically with the
recognition by Abbe (1967) that the Large and Small Magellanic Clouds
(\object{LMC} \& \object{SMC}) are in fact nearby galaxies composed of stars
and nebular matter, separate from our own \object{Milky Way} galaxy. Four
decades later, the discovery in 1907 by Henrietta Leavitt of a relation
between the pulsation period and apparent brightness of Cepheid-type variables
in the \object{LMC} counts amongst the most important discoveries in modern
astronomy.

Over the past few decades, a great deal of astro-physics has been learnt from
studies of stars in the Magellanic Clouds, for several reasons: (i) stars in
each of the Clouds are all at the same distance to us, and therefore their
relative luminosities are well determined; (ii) this distance is well known;
(iii) the Clouds are near enough to study individual stars; (iv) interstellar
extinction to and within the Clouds is small; (v) as dwarf galaxies, the
Clouds are fairly big, and they have experienced star formation over much of
their history. Hence, plenty of stars can be found in different evolutionary
stages, including stars in clusters for which accurate ages may be determined.
(vi) The \object{LMC} and \object{SMC} have metallicities of $\sim\frac{1}{2}$
and $\frac{1}{5}$ solar, allowing the study of metallicity effects. However,
as the analyses and astrophysical problems they address become more detailed,
these advantages meet their limitations: (i) the depth of the Clouds has been
measured and is not always negligible; (ii) different methods yield a distance
modulus of the \object{LMC} ranging from $\sim18.1$ to 18.7 mag (I adopt
distance moduli to the \object{LMC} and \object{SMC} of 18.5 and 19.0 mag,
respectively); (iii) current instrumentation does not always provide the
required spatial resolution or detection sensitivity; (iv) although extinction
is much less than in the plane of the \object{Milky Way}, it is not always
negligible even at infrared (IR) wavelengths, and can be very patchy in
places; (v) the number of stars in very rare, short-lived evolutionary phases
is limited --- sometimes to less than one at any moment --- hampering
statistical studies. Also, the star formation history of the \object{LMC} is
complex, with a lack of 4 to 10 Gyr-old stellar clusters. And (vi) there are
no (identified) samples of either metal-rich stars, with [Fe/H]$>0$, or
extremely metal-poor stars, with [Fe/H]$\ll-2$.

\section{Asymptotic Giant Branch (AGB) stars}

The AGB phase is the final stadium in the life of a star with a Zero-Age
Main-Sequence (ZAMS) mass of $\sim0.5$ to 8 M$_\odot$. The lower limit to this
mass range is set by the requirement that the temperature and pressure at the
bottom of the hydrogen mantle surrounding the electron-degenerate core (oxygen
and carbon/neon) is sufficiently high to ignite hydrogen shell burning,
driving the evolution of the star along the asymptotic giant branch towards
higher luminosity (growing core) and lower surface temperature (expanding
mantle). The AGB evolution terminates either because the core has reached the
Chandrasekhar mass limit of $M_{\rm max}\sim1.4$ M$_\odot$ beyond which the
core becomes gravitationally unstable, the corresponding luminosity of $L_{\rm
max}\sim55,000$ L$_\odot$ defining the tip of the AGB, or because the mantle
has diminished to the extent that the hydrogen-burning shell becomes exposed
and quickly runs out of fuel, leaving the hot and compact core (white dwarf)
to cool into oblivion. The upper limit to the mass range of AGB progenitors is
set by the condition for the ignition of carbon burning in the non-degenerate
core of supergiants.

The hydrogen-burning shell produces an underlying helium shell which ignites
as soon as sufficient pressure has built up, thereby causing the mantle to
expand and the hydrogen burning to switch off. Such Thermal Pulses (TP)
re-occur on timescales from ${\Delta}t\sim10^3$ yr for the most massive AGB
stars up to ${\Delta}t\sim10^5$ yr for the least massive AGB stars. The TP
itself is brief ($\sim1$\% of ${\Delta}t$), but it takes long (up to
$\sim30$\% of ${\Delta}t$) for a star to recover from the luminosity dip of up
to a factor two whilst the hydrogen shell burning gradually takes over from
the helium shell burning. The effect of the TP on the mantle and emergent
luminosity is relatively stronger for less massive stars. The TP-AGB phase
lasts for $t=$few$\times10^5$ yr, up to $t=$few$\times10^6$ yr for the most
massive AGB stars unless mass loss is severe.

Energy transport from the nuclear burning site to the stellar surface of an
AGB star proceeds via convective motion through the mantle: a rising gas
element continues to rise. Much theoretical understanding needs to be gained
about the exact criterium for convective instability and the description of
convective flows and boundaries, especially because convection not only
transports energy but also matter and hence affects the stratification of
elemental abundances throughout the mantle. When during a TP the convection
zone penetrates into layers containing the products of nuclear burning, these
heavy elements will become mixed into and become over-abundant in the stellar
photosphere where light leaves the star. This is the third time in the
evolution of intermediate-mass stars that such dredge-up event may happen, and
probably occurs only for the more massive AGB stars ($M\ga1.5$ M$_\odot$).
Products of the slow neutron-capture process, s-process elements such as
zirconium, technetium and barium are used to diagnose the TP-AGB history of
stars. Carbon may become enhanced too, greatly affecting the chemistry in the
cool stellar atmosphere ($T_{\rm eff}\la4000$ K) if after binding carbon and
oxygen into carbon-monoxide (CO) not oxygen but carbon remains to form more
complex molecules. Such carbon stars are easily distinguished from oxygen-rich
AGB stars as their spectra are dominated by absorption bands of CN, HCN, C$_2$
and C$_2$H$_2$ rather than TiO, VO and SiO. However, nuclear burning may take
place at the inner boundary of the convection zone, this Hot Bottom Burning
(HBB) converting carbon into oxygen and nitrogen and preventing the formation
of a carbon star. HBB occurs in massive AGB stars ($M\ga4$ M$_\odot$) and
enhances their luminosity, but switches off once the mantle has nearly gone
due to severe mass loss.

The stability criterium for motion within the mantle not only determines the
occurrence of convection (turbulent motion), but also the occurrence of
pulsation (organised motion). In the hydrogen ionization zone not far below
the stellar surface, and in the first helium ionization zone, the absorption
coefficient of the gas to radiation from the stellar interior is very large;
hence the internal energy becomes high and the mantle expands until the
opacity drops and the mantle shrinks again. These radial pulsations levitate
the atmosphere high enough for dust formation to occur and radiation pressure
to blow away the circumstellar dust --- taking with it the collisionally
coupled gas. Mass-loss rates of $\dot{M}\sim10^{-7}$ to $10^{-4}$ M$_\odot$
yr$^{-1}$ cause the pre-mature termination of AGB evolution, exposing the
white dwarf after the star has shed its mantle.

\section{AGB stars in the Magellanic Clouds}

Searches for AGB stars in the Magellanic Clouds have followed three main
strategies: (i) monitoring surveys to find Long-Period Variables (LPVs),
mostly of \object{Mira} type defined by regular pulsation on the order of a
year with a visual amplitude $A_{\rm V}\geq2.5$ mag; (ii) spectroscopic or
photometric surveys to find bright M-type or carbon stars, and (iii) mid-IR
surveys to find AGB stars with circumstellar dust emission.

\subsection{Optically bright AGB stars}

Amongst the most influential systematic searches for magellanic M-type giants
and carbon stars are the study of selected fields in the bar of the
\object{LMC} and in the \object{SMC} (Blanco et al.\ 1980); Blanco \& McCarthy
1983; Frogel \& Richer 1983), the construction of the carbon star luminosity
function in the \object{LMC} (Costa \& Frogel 1996), and the survey of
magellanic clusters for which ages and hence the ZAMS masses of the cluster
members can be determined (Aaronson \& Mould 1985; Frogel et al.\ 1990;
Westerlund et al.\ 1991).

\begin{figure}
\begin{center}
\mbox{
\hspace*{-5mm}
\epsfxsize=0.54\textwidth\epsfysize=0.54\textwidth\epsfbox{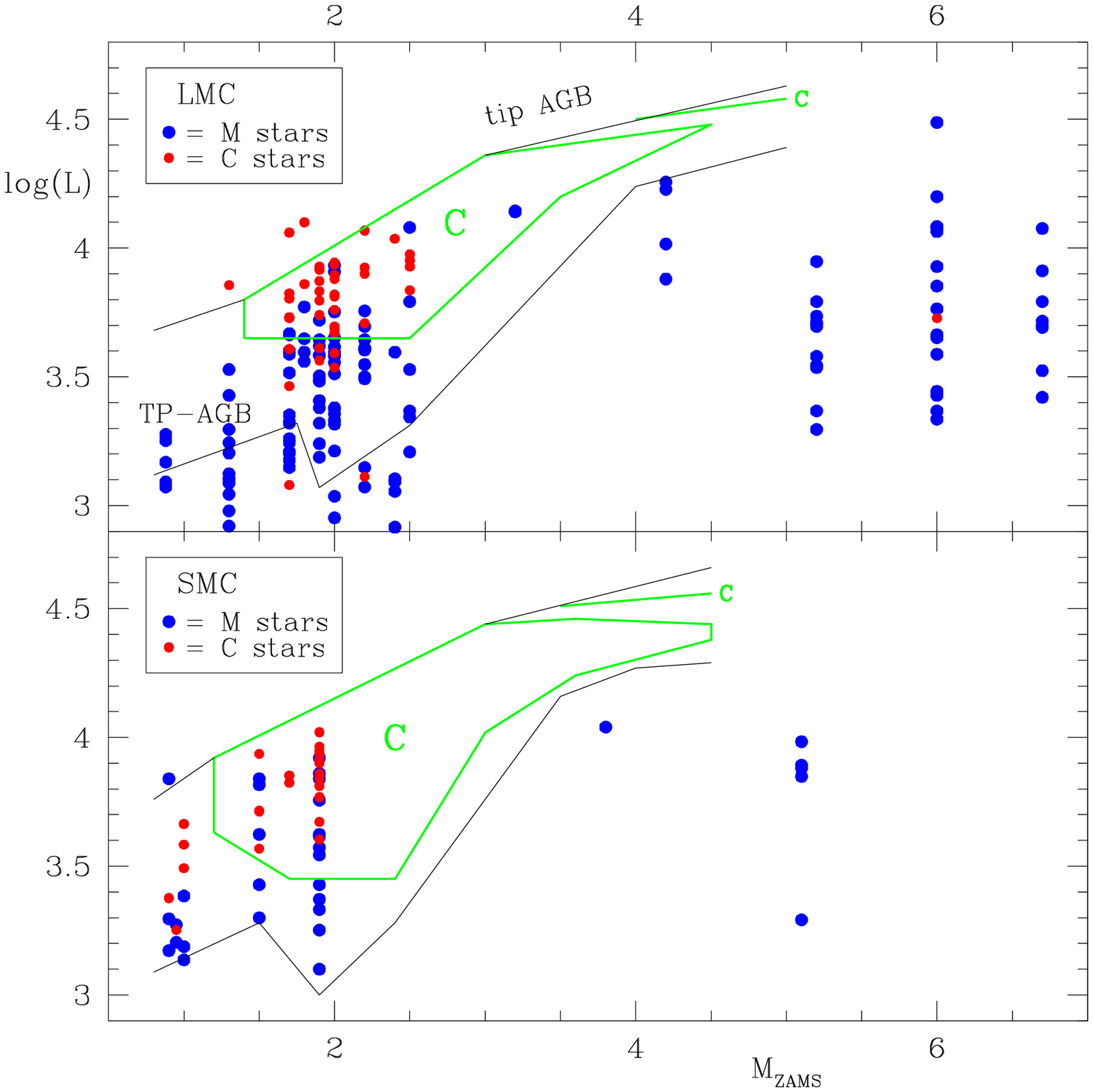}
\hspace*{-7mm}
\epsfxsize=0.54\textwidth\epsfysize=0.54\textwidth\epsfbox{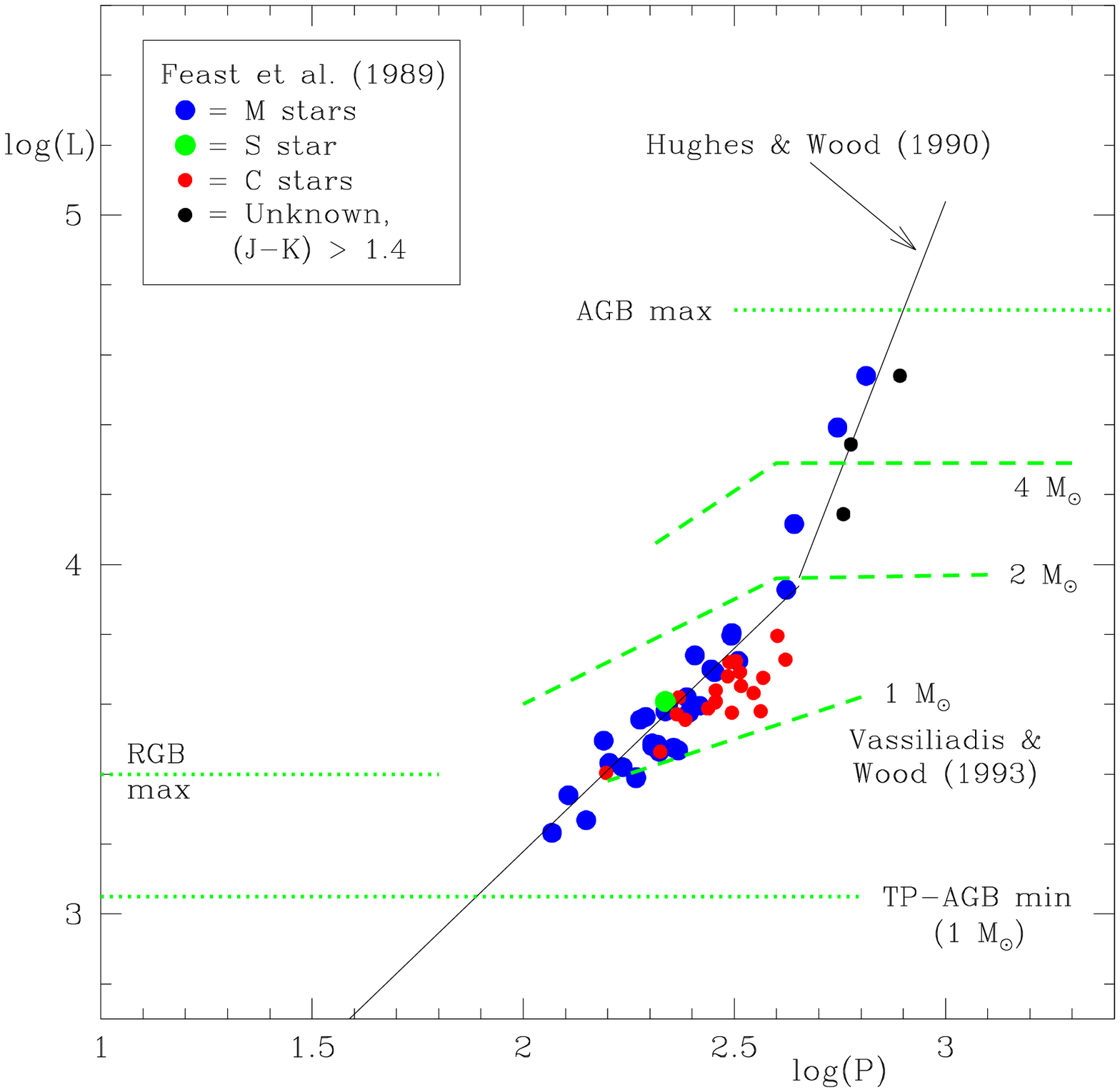}
}
\caption{{\bf Left:} models (Marigo et al.\ 1999) and cluster data (Frogel et
al.\ 1990); {\bf Right:} P-L relation in the \object{LMC} (Feast et al.\ 1989;
Hughes \& Wood 1990) and evolutionary tracks (Vassiliadis \& Wood 1993).}
\end{center}
\end{figure}

Although much of the physics remains in parameterised form --- notably
convection and dredge-up --- evolutionary models can be tuned to reproduce the
carbon star luminosity function quite well (Marigo et al.\ 1999). When these
results are confronted with the luminosities and chemistry of cluster AGB
stars in the Magellanic Clouds (Frogel et al.\ 1990), the match is not
entirely satisfactory (Fig.\ 1 left) due to both imperfections in the models
as well as difficulties in deducing the observed quantities.

The first grand survey for variable stars resulted in the catalogue of Harvard
Variables (Payne-Gaposchkin 1971) which, however, were mostly supergiants.
Subsequent surveys identified pulsating AGB stars in sufficient numbers for a
P-L relation to be proposed (Glass \& Lloyd-Evans 1981) and refined (Wood et
al.\ 1983; Feast et al.\ 1989; Hughes \& Wood 1990). Mira's did indeed seem to
obey a remarkably tight and unique P-L relation (Fig.\ 1 right) suggesting a
robust instability threshold and a single pulsation mode, and their possible
use as distance indicators, but already a few complications may be noticed: (i)
The P-L relation has a different slope for periods shorter/longer than
$P^\prime\sim400$ d, which evolutionary tracks (Vassiliadis \& Wood 1993) link
to a ZAMS mass $M^\prime\sim2$ M$_\odot$. Previous evolution, for instance
whether ($M<M^\prime$) or not ($M>M^\prime$) the star experienced violent
ignition of core-helium burning at the tip of the first red giant branch
(RGB), may be the cause for a difference in internal structure and pulsational
properties. Also, (ii) carbon stars and especially red stars have somewhat
larger periods for their luminosities, which may be due to their larger radii
as they have suffered significant mass loss. (iii) A paucity of luminous AGB
stars, in particular carbon stars with $L\ga20,000$ L$_\odot$ corresponding to
$M\ga4$ M$_\odot$, was explained as due to HBB (Iben 1981; Wood et al.\ 1983).
This was confirmed by the identification of lithium-enriched S-type stars (C/O
just below unity) with $20,000<L<55,000$ L$_\odot$ (Smith et al.\ 1995).

\subsection{Dust-enshrouded AGB stars}

Studies of optically bright AGB stars are biased against stars in the final
phases of AGB evolution that have become dust-enshrouded as a result of mass
loss. It is difficult to study their properties as their stellar surface is
obscured at optical wavelengths and the photosphere at IR wavelengths becomes
dominated by thermal emission from the circumstellar dust.

The IRAS mission in the early 1980s provided a catalogue with a few hundred
mid-IR point sources in the Magellanic Clouds that are candidate dust shells
around AGB stars (Loup et al.\ 1997). Follow-up on such IRAS sources and
similar dust-enshrouded supergiants (Wood et al.\ 1992; van Loon et al.\ 1998,
and references therein) culminated in the detailed quantitative study of
$\sim50$ dust-enshrouded AGB stars in the \object{LMC} observed with ISO (van
Loon et al.\ 1999). The main results are that (i) luminous carbon stars do
exist amongst dust-enshrouded AGB stars, confirming the HBB scenario with mass
loss; (ii) the threshold above which HBB occurs is confirmed to be 4 M$_\odot$
(van Loon et al.\ 2001); (iii) winds of AGB stars are dust-driven (e.g.\ Gail
\& Sedlmayr 1986); (iv) mass-loss rates exceed nuclear burning mass
consumption rates and reach a maximum $\dot{M}_{\rm max}{\propto}L$ in a
distinct superwind phase; (iv) mass-loss rates increase with lower $T_{\rm
eff}$ --- this effect is strongest during the early stages of circumstellar
dust formation --- but may be only little dependent on metallicity for
$[Fe/H]\ga-1$ (van Loon 2000). Great improvement in our knowledge of AGB mass
loss is expected from recent surveys of the Magellanic Clouds at near-IR
(2MASS, DENIS) and mid-IR (MSX, ISO) wavelengths, and their follow-up.

\subsection{Large-scale variability surveys}

\begin{figure}
\begin{center}
\mbox{
\hspace*{-6mm}
\epsfxsize=0.55\textwidth\epsfysize=0.55\textwidth\epsfbox{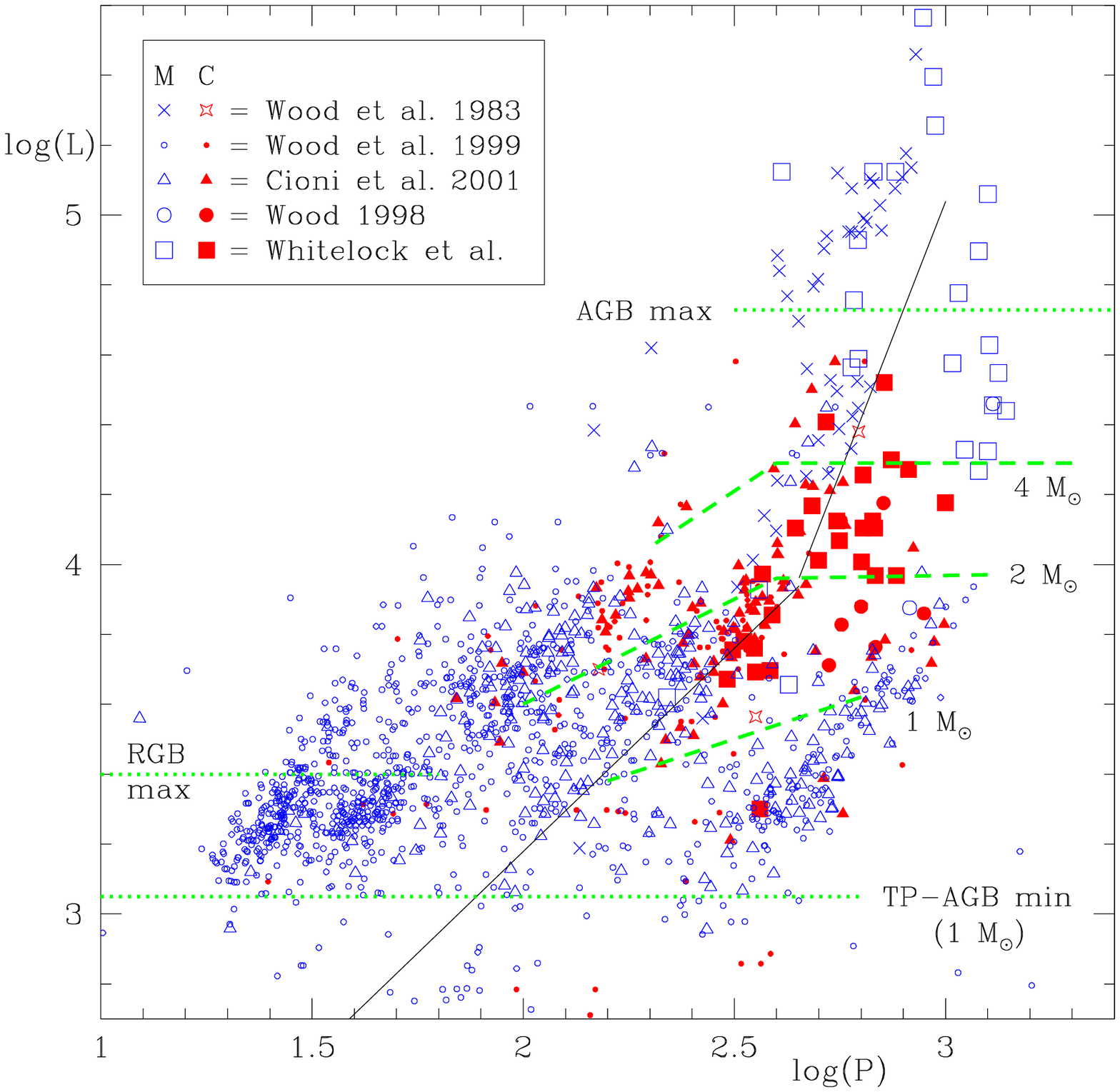}
\hspace*{-8mm}
\epsfxsize=0.55\textwidth\epsfysize=0.55\textwidth\epsfbox{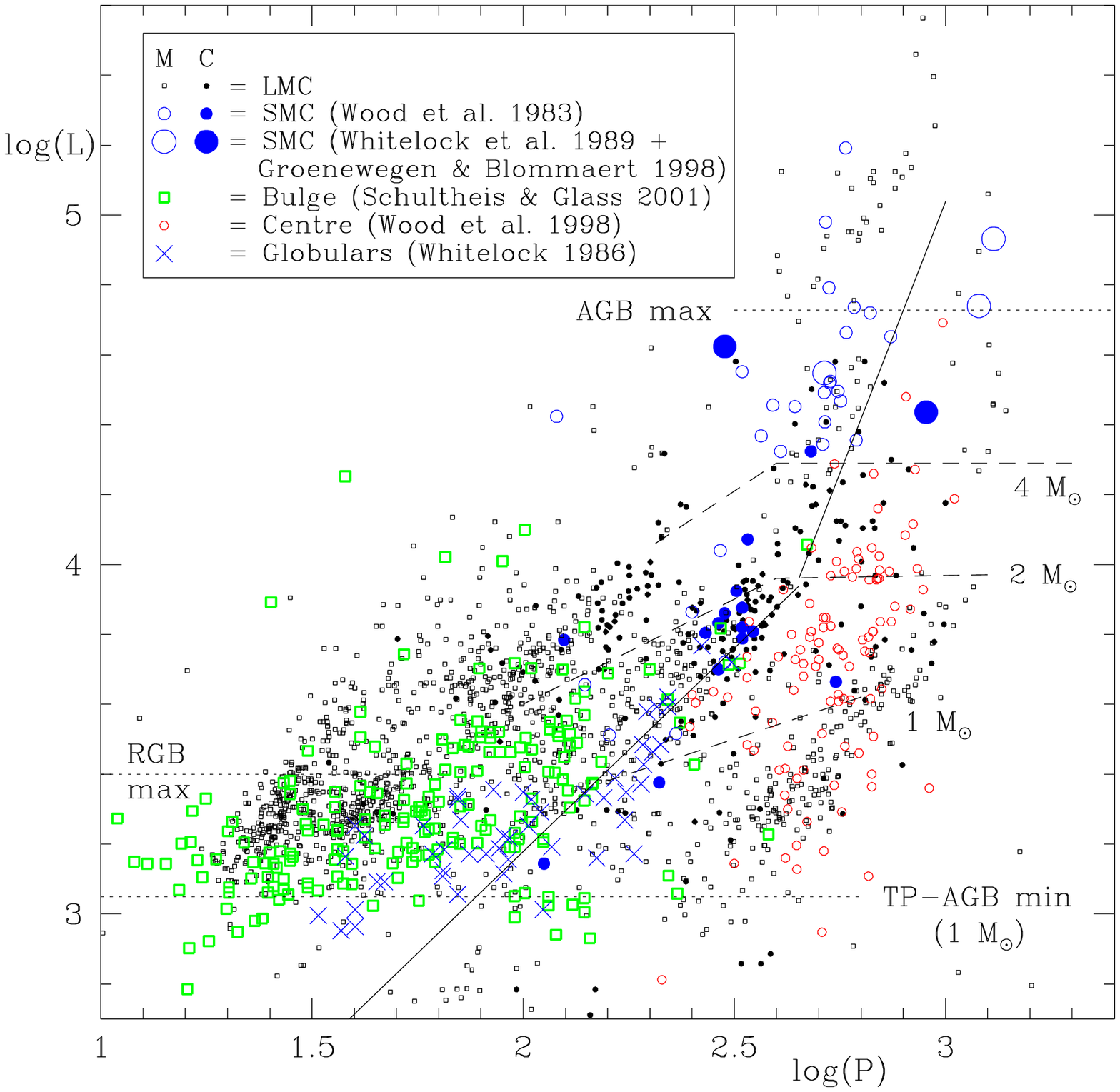}
}
\caption{P-L diagram in the \object{LMC} ({\bf left}), and compared with the
\object{SMC} and the galactic bulge, centre and globular clusters ({\bf
right}).}
\end{center}
\end{figure}

Much progress is made in detecting variables in the \object{LMC} thanks to
micro-lensing surveys (MACHO, EROS, OGLE). The P-L diagram (Fig.\ 2, left) not
only shows the classical \object{Mira} sequence but at least three more
roughly parallel sequences. Two of these, at shorter periods and populated by
semi-regulars (many of which are regular but have smaller amplitudes than
Miras), are believed to correspond to overtone modes with the Miras pulsating
in the fundamental mode (Wood \& Sebo 1996; Wood et al.\ 1999). This
interpretation is supported by the behaviour of pulsation amplitudes \&
velocities (Scholz \& Wood 2000), period \& radius, and non-linear models
(Ya'ari \& Tuchman 1996, 1999) for galactic Miras, but if the mixing length
describing convection is allowed to vary then Miras could still pulsate in the
first overtone (Barth\`{e}s 1998). Another sequence at longer periods than the
Mira P-L is explained as binarity (Wood et al.\ 1999).

It is estimated that not all but $\sim65$\% of the TP-AGB stars are LPV, of
which (only) $\sim10$\% are Miras (Cioni et al.\ 2001), and the evolution in
the P-L plane is complex: the pulsation is a function of mass and time --- not
necessarily in a continuous or even monotonous fashion. It is then remarkable
how consistent the P-L diagram is populated by stars in the \object{LMC},
\object{SMC} and galactic centre, and the galactic bulge ($M\la2$ M$_\odot$)
and globular clusters ($M\la1$ M$_\odot$) despite an order of magnitude
difference in metallicity (Fig.\ 2, right).

An overdensity of stars at the two short-period sequences just below the
maximum luminosity at the RGB indicates these may be RGB stars, whilst above
the maximum luminosity at the AGB, supergiants are found on sequences similar
to but slightly displaced from the Mira P-L and the long-period sequence. This
suggests that certain similarities in the internal structure of RGB stars, AGB
stars and (red) supergiants cause similarities in their pulsation.

Carbon stars populate the Mira P-L where stars with $M_{\rm ZAMS}\sim1.5$ to 4
M$_\odot$ are found, as expected from third dredge-up and HBB. Carbon stars on
the sequence immediately at shorter periods than the \object{Mira} P-L, at
$L\sim8,000$ L$_\odot$, may have $M_{\rm ZAMS}\sim3$ to 4 M$_\odot$, but the
continuation of the evolutionary track is devoid of stars until it reaches the
\object{Mira} P-L at a twice as high luminosity. Perhaps, after having become
carbon stars on the \object{Mira} P-L, these stars are temporarily
semi-regulars, possibly due to a TP. Surprisingly, the most luminous carbon
stars obey the \object{Mira} P-L. Thought to have become carbon-enriched after
an extended period of severe mass loss terminated HBB, they would be expected
to have the longest periods of all --- longer than on the \object{Mira} P-L.

\section{The connection between pulsation and mass loss of AGB stars}

When AGB stars arrive on the \object{Mira} P-L, their mass-loss rates have
become so severe that further evolution is along tracks of constant
luminosity, with the radius (and hence pulsation period) growing as the
structure of the mantle re-adjusts itself to the decrease in mass. Hence the
dust-enshrouded stars (Wood 1998; Whitelock et al., in preparation) are on the
\object{Mira} P-L or to longer periods. Describing the \object{Mira} as a
polytrope ($R\sqrt[3]{M}=constant$) and the pulsation as a harmonic oscillator
($2\pi/P \simeq 1/t_{\rm freefall} \simeq g(R)/R$), the period evolves as
$P{\propto}1/M$. For instance, a \object{Mira} with $P=250$ d ($M_{\rm
ZAMS}=1.5$ M$_\odot$) would end up on the long-period sequence at $P\sim750$ d
and $M\sim0.5$ M$_\odot$, which is a reasonable white dwarf mass. The longest
observed period for an evolved equivalent of a \object{Mira} with $P=700$ d
($M_{\rm ZAMS}=5$ M$_\odot$), however, has $P\sim1400$ d and $M\sim2.5$
M$_\odot$. Such star must still lose more than a solar mass, but stars with
correspondingly long periods ($P\sim3000$ d) have not (yet) been found.

The growth in radius/period of mass-losing AGB stars may become truncated when
the dynamical timescale, $P$, approaches the thermal timescale,
$t{\simeq}GM^2/LR$, and the mantle no longer pulsates adiabatically. This
happens when $P{\simeq}\sqrt[3]{{\pi}M/{\sigma}T_{\rm eff}^4}$, corresponding
to $P\sim2000$ d for a star of $M=2.5$ M$_\odot$ and $T_{\rm eff}=2500$ K
(note the moderate dependence on stellar parameters). By this time, the
surface gravity is so low that mass loss could be initiated without the help
of pulsation: the star may quite literally be boiling over, and this could be
responsible for the superwind phase required to shed the remainder of the
mantle that may become visible as a planetary nebula. The ability of the
mantle to restore thermo-dynamic equilibrium on a dynamical timescale may also
be relevant for the faint part of the long-period sequence ($L\la10,000$
L$_\odot$). Some disturbance, for instance tidal interaction in a binary,
could trigger instabilities on a thermal (not orbital) timescale, causing
irregular mass loss that results in obscuration events, super-imposed on a
``traditional'' (adiabatic) pulsation in the fundamental mode (\object{Mira}
P-L) or overtone mode (shorter-period sequences).

\begin{figure}
\begin{center}
\mbox{
\hspace*{-7mm}
\epsfxsize=0.55\textwidth\epsfysize=0.55\textwidth\epsfbox{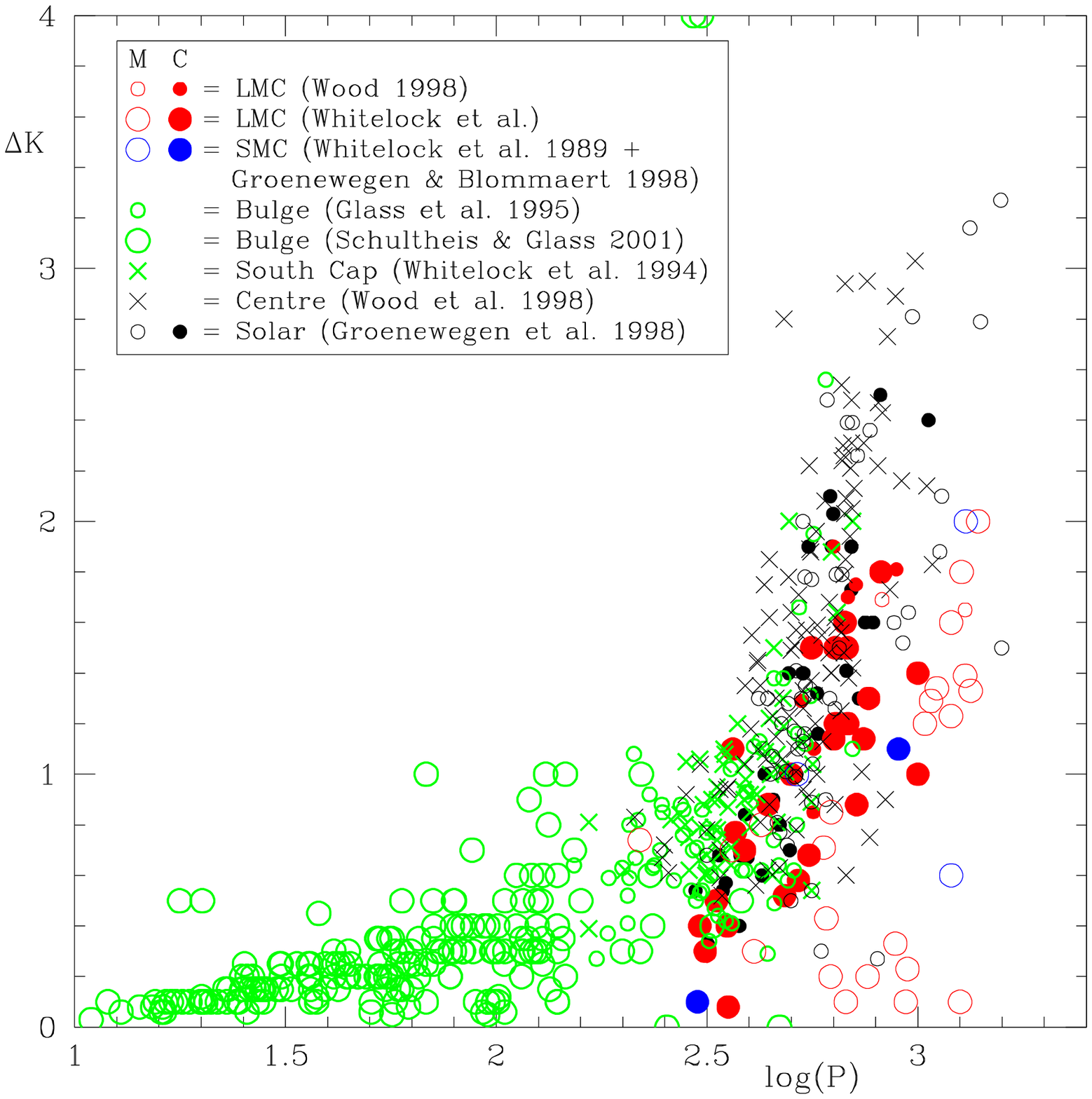}
\hspace*{-7mm}
\epsfxsize=0.55\textwidth\epsfysize=0.55\textwidth\epsfbox{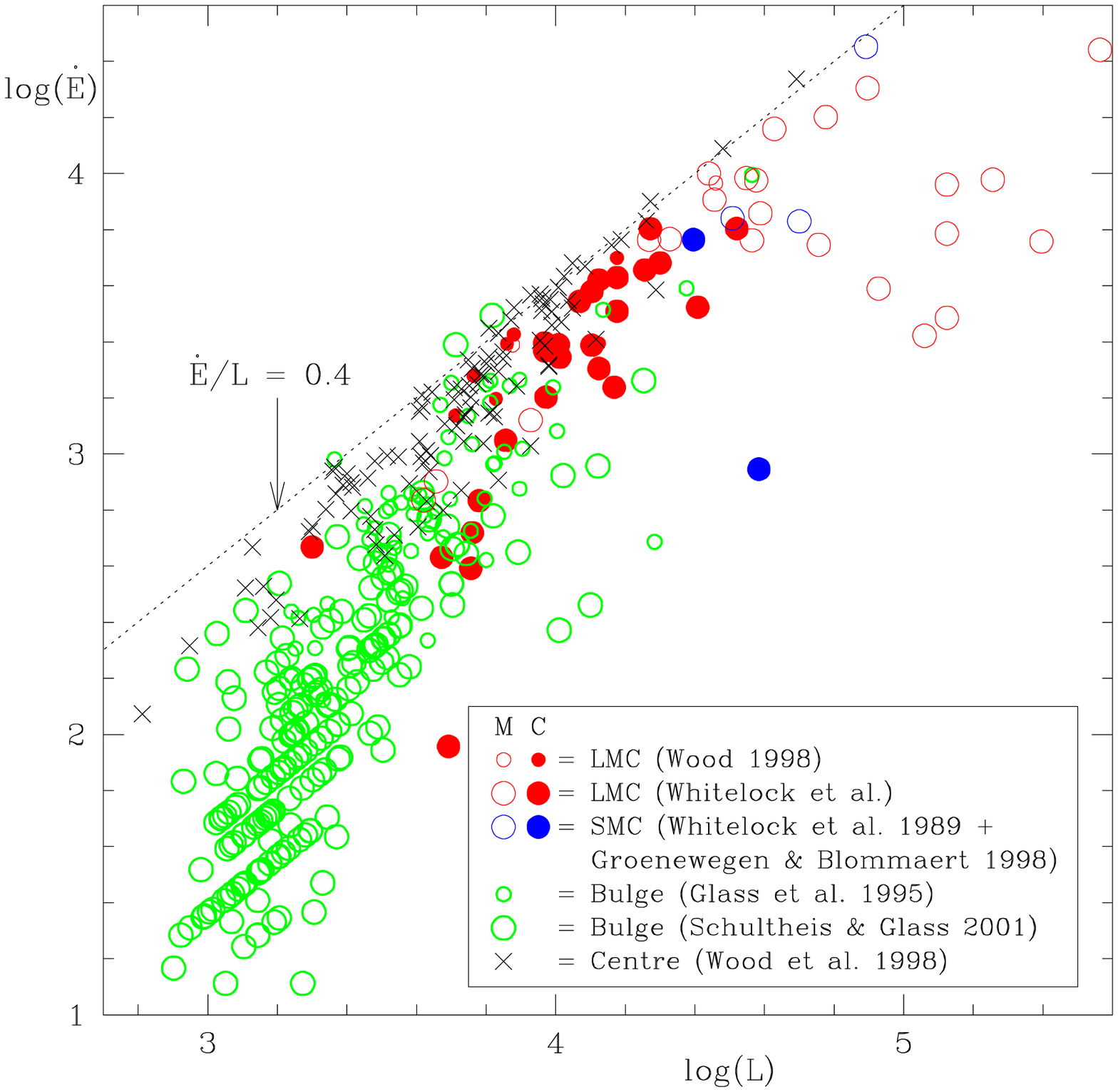}
}
\caption{K-band amplitudes and pulsation energy input rate, versus period
({\bf left}) and luminosity ({\bf right}) in the \object{LMC}, \object{SMC}
and \object{Milky Way}.}
\end{center}
\end{figure}

K-band amplitudes (a measure for the bolometric amplitude) generally increase
with period, but for $P\ga400$ d the increase becomes steeper (Fig.\ 3, left).
This could be related to the different slope of the \object{Mira} P-L relation
for periods shorter/longer than $P^\prime\sim400$ d. The diagram suggests
that, for longer periods, carbon stars and stars in the Magellanic Clouds
pulsate less vigorously than galactic oxygen-rich stars. However, the
pulsation of a luminous star with small amplitude in magnitude may well
involve more energy than that of a faint star with large amplitude. The
time-averaged rate at which radiative energy is stored in the mechanical
pulsation, $\dot{E}=L/2\times(\exp[{\Delta}K/2.5]-1)/(\exp[{\Delta}K/2.5]+1)$
has a theoretical maximum of $\dot{E}_{\rm max}/L=0.5$: the maximum energy
rate available during that half of the pulsation cycle in which the mantle is
lifted is $L$. The observed $\dot{E}$ increases with luminosity (Fig.\ 3,
right), saturating at $\dot{E}_{\rm max}/L=0.4$ --- i.e.\ very efficient ---
with no clear distinction between stars of different chemistry. As the maximum
mass-loss rate was found to increase proportionally to luminosity (van Loon et
al.\ 1999), I obtain $\dot{M}_{\rm max}\propto\dot{E}_{\rm max}$. This
explains the absence of a clear metallicity-dependence of the mass-loss rate
(van Loon 2000).

\end{document}